\documentclass[useAMS,usenatbib]{mn2e}

\usepackage{graphicx}
\usepackage{amsmath}
\usepackage{amssymb}
\usepackage{color}
\usepackage{subfig}

\title[Globular Clusters and the Gamma-Ray Excess in the GC]{Disrupted Globular Clusters and the Gamma-Ray Excess in the Galactic Centre}
\author[Giacomo Fragione, Fabio Antonini, Oleg Y. Gnedin]{Giacomo Fragione$^{1}$\thanks{E-mail: giacomo.fragione@mail.huji.ac.il}, Fabio Antonini$^{2}\thanks{E-mail: fabio.antonini@northwestern.edu}$ and Oleg~Y. Gnedin$^{3}$\\
$^{1}$Racah Institute for Physics, The Hebrew University, Jerusalem 91904, Israel\\
$^{2}$Center for Interdisciplinary Exploration and Research in Astrophysics (CIERA), Northwestern University, Evanston, IL 60208, USA\\
$^{3}$University of Michigan, Department of Astronomy, Ann Arbor, MI 48109, USA}

\begin{document}

\maketitle

\begin{abstract}
The Fermi Large Area Telescope has provided the most detailed view toward the Galactic Centre (GC) in high-energy gamma rays. Besides the interstellar emission and point-source contributions, the data suggest a residual diffuse gamma-ray excess. The similarity of its spatial distribution with the expected profile of dark matter has led to claims that this may be evidence for dark matter particle annihilation. Here, we investigate an alternative explanation that the signal originates from millisecond pulsars (MSPs) formed in dense globular clusters and deposited at the GC as a consequence of cluster inspiral and tidal disruption. We use a semi-analytical model to calculate the formation, migration, and disruption of globular clusters in the Galaxy. Our model reproduces the mass of the nuclear star cluster and the present-day radial and mass distribution of globular clusters. For the first time, we calculate the evolution of MSPs from disrupted globular clusters throughout the age of the Galaxy and consistently include the effect of the MSP spin-down due to magnetic-dipole breaking. The final gamma-ray amplitude and spatial distribution are in good agreement with the Fermi observations and provide a natural astrophysical explanation for the GC excess.
\end{abstract}


\section{Introduction}

The Large Area Telescope instrument on board the Fermi Gamma-Ray Space Telescope (Fermi-LAT) has provided high-quality data in the energy range from $20$~MeV to over $300$~GeV. Several groups analysed the data toward the region surrounding the Galactic Centre (GC) \citep{gor13,aba14,cal15,cac15,pet15,aje16}. Such analyses found a gamma-ray excess around the GC, peaking at $\approx 2$~GeV, with approximately spherical density profile $\propto r^{-2.4}$, out to $3\,$kpc from the GC. Due to the similarity with the predicted radial distribution of dark matter \citep[e.g.,][]{nfw97}, dark matter annihilation has been suggested as a possible explanation for the GC excess. However, this interpretation is challenged by non-detection of the corresponding signal from dwarf spheroidal satellite galaxies of the Milky Way \citep{alb17}. These galaxies are strongly dark-matter dominated and should produce a detectable signal if dark-matter annihilation was indeed the cause of the GC excess. In addition, \citet{aje17} have recently shown that the dark matter interpretation cannot account for the observed properties of the population of pulsar candidates in the \textit{Fermi} Collaboration's analysis.

An alternative explanation is the emission of thousands of unresolved milli-second pulsars (MSPs) \citep{bra15,bar16,abb17,arc17,aje17}, which have very similar gamma-ray spectral energy distribution to that of the GC excess \citep{aba14,bra15}. Recently, \citet{lee15} and \citet{bar16} showed evidence of the existence of an unresolved MSP population in the inner $20^\circ$ of the GC, with the spatial distribution and total flux consistent with the \textit{Fermi} data. MSPs are thought to be related to low-mass X-ray binary (LMXB) systems. X-rays in these systems are emitted by an accretion disk of a neutron star, which is stripping the outer layers of the companion star. The resulting transfer of angular momentum decreases the rotation period of the pulsar to milliseconds. Such binary systems form in dense environments where the rate of stellar dynamical encounters is high. \textit{In situ} formation of MSPs at the GC is disfavoured since LMXBs are quite rare in the bulge of our Galaxy \citep{cho15,hag17}. \citet{bra15}, following an idea first proposed by \citet{aba12} and \citet{bed13}, suggested instead that the excess may be explained by the population of MSPs left in the inner region of the Milky Way as a consequence of globular cluster disruption \citep{ant12,ant13,fgk17}. In this model, the MSPs are formed in globular clusters, which migrate inward through dynamical friction and tidally dissolve in the GC, leaving behind a population of MSPs and LMXBs. While the population of LMXBs is expected to exhaust its energy and become too faint to be detected after $\sim 10^{8}\,$yr, the MSP population would last much longer due to the long timescales needed for significant spin-down. This is consistent with the few ($42$) LMXBs observed by \textit{INTEGRAL} within a $10^\circ$ radius around the GC \citep{hag17}.

Does the MSP explanation work in detail? \citet{bra15} argued that it results in the amplitude, angular distribution, and spectral signatures of the gamma-ray excess that are in good agreement with observations. They assumed the same gamma-ray luminosity per unit mass for the disrupted clusters as for present-day clusters, but did not include the spin-down of MSPs as a consequence of magnetic-dipole braking. \citet{hol16} suggested that, if the MSP spin-down was taken into account, such an astrophysical explanation could account for only a small fraction of the excess, while \citet{pet15} pointed out the importance of details of the MSP luminosity function and the secondary emission via inverse Compton losses of electrons injected in the interstellar medium. Moreover, \citet{hom16} and \citet{hol16} argued that in this scenario \textit{Fermi} should have spotted tens of point sources within $\approx 10^\circ$ of the GC. In contrast, \citet{aje17} showed that a model that assumes $\approx 2.7$ pulsars in the Galactic disk for each pulsar in the bulge is consistent with the population of known gamma-ray pulsars as well as with the spatial profile and energy spectrum of the GC excess. Thus there is still no consensus on the origin of the gamma-ray excess.

In this paper, we reconsider the MSP scenario in greater detail. We model the formation and disruption of Galactic globular clusters across all cosmic time and calculate the amount of MSPs deposited in the Galactic bulge as a consequence of cluster disruption. Unlike \citet{bra15}, we take into account the merger history of globular clusters across the age of the Milky Way and the spin-down of the MSPs due to magnetic-dipole braking self-consistently. Unlike \citet{hol16}, we adopt a spin-down distribution that is more consistent with observational constraints \citep{fre01a,fre01b,pra17}. This allows us to estimate the present day gamma-ray luminosity from the bulge more accurately and compare it with the observed GC excess more consistently.

Our paper is organized as follows. In Section 2, we describe our semi-analytical model to create and evolve the population of Galactic globular clusters. We also present our model to evolve MSP luminosities and our choice of the MSP spin-down timescale distribution. In Section 3, we show that our results agree with the most recent measurements of the GC excess. In Section 4, we discuss our results and compare them to previous works. In Section 5, we summarize our conclusions.

\section{Model for the gamma-ray luminosity of milli-second pulsars}
\label{sec:model}

In this section, we describe the equations used to estimate the present gamma-ray luminosity generated by the emission of MSPs. First, we describe our model for the formation and dynamical evolution of globular clusters across the age of the Galaxy, including stellar mass loss, tidal stripping of stars, and tidal disruption. Then we introduce the equations that we used to compute the evolution and spin-down of the MSP population, and describe the observational constraints on the distribution of spin-down period. Finally, we derive a log-linear relation between the mass of globular clusters and their gamma-ray luminosity.

\subsection{Globular cluster formation and evolution}

\begin{figure*} 
\centering
\begin{minipage}{18cm}
\subfloat{\includegraphics[scale=0.55]{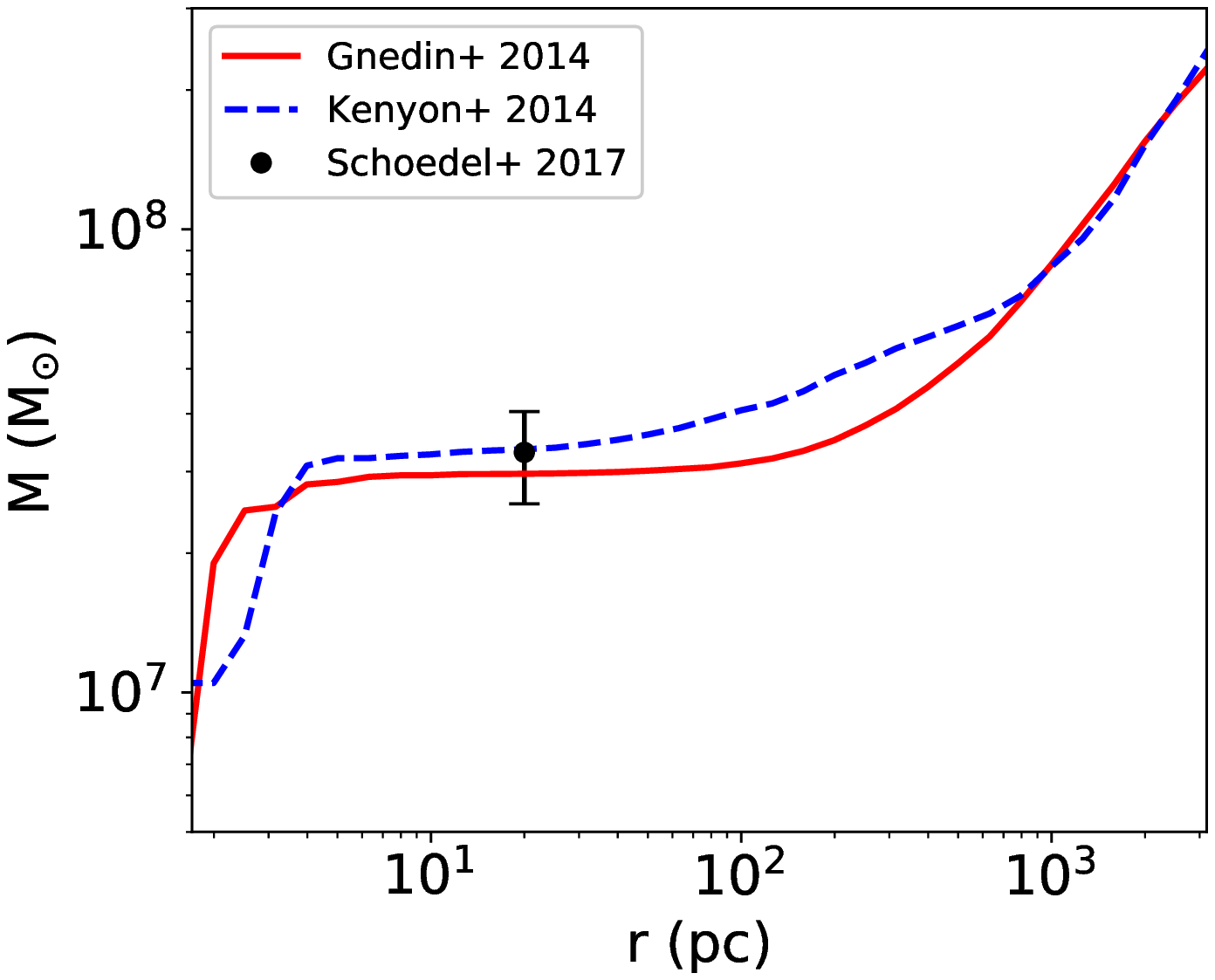}\label{fig:mldep}}
\hspace{1cm}
\subfloat{\includegraphics[scale=0.55]{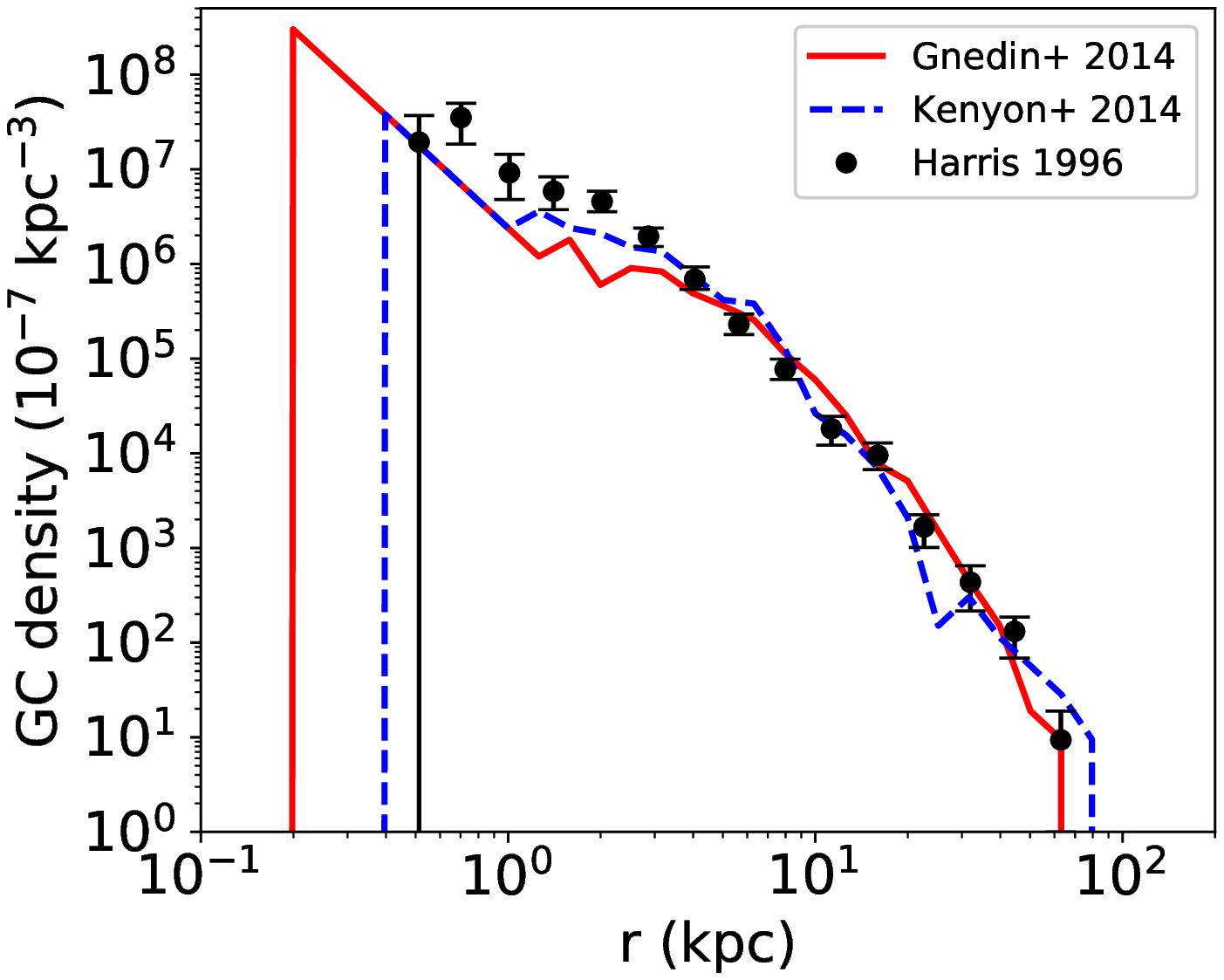}\label{fig:gcdens}}
\caption{\textit{Left:} Cumulative stellar mass deposited by disrupted globular clusters as a function of Galactocentric radius for \citet{gne14} and \citet{ken14} potential models. The black point at $3.3\times 10^7\,\mathrm{M}_{\odot}$ is the stellar mass of the nuclear cluster inside $20\,$pc as measured by \citet{sch17}. \textit{Right:} Number density of Galactic globular clusters surviving to the present time for \citet{gne14} and \citet{ken14} potential models, compared with the number density of observed Milky Way's globular clusters \citep[2010 version of][]{har96}.}\label{fig1}
\end{minipage}
\end{figure*}

We use the \citet{gne14} model to evolve the Milky Way's globular cluster population. Here we briefly discuss the equations used to calculate the model. Since the total mass of the globular cluster system is roughly proportional to the total stellar mass in massive galaxies at the present time, we assume that the cluster formation rate was a fixed fraction $f_{GC,i}$ of the overall star formation rate
\begin{equation}
  \frac{dM_{GC}}{dt} = f_{GC,i}\, \frac{dM_{*}}{dt}\ .
\end{equation}
We set $f_{GC,i}=0.011$. This choice leads to a mass of $3.3\times 10^7\,\mathrm{M}_{\odot}$ deposited within $20$ pc of the GC by the present time. This is consistent with the observed mass of the nuclear star cluster, inferred by \citet{sch17} (see left panel of Figure~\ref{fig1}). Thus, we work under the assumption that a large fraction of the Milky Way mass
inside $\sim 100\rm pc$ was brought in by infalling globular clusters \citep{ant12}. 
While observations show that in-situ star formation also contributed to the build up of
the stellar populations inside $\sim 1 \rm pc$
\citep[e.g.,][]{fel17}, the
relative  contribution of the two formation channels (in-situ vs cluster inspirals) remains an open question.
Moreover, we assume that the initial density distribution of clusters is proportional to the density distribution of field stars formed at the same epoch. Hence we initialize the globular clusters in a spherical distribution, proportional to that of the field stars, and treat their position as the time-averaged radius of an assumed orbit, as in \citet{gne14}. As discuss later in \S\ref{sec:pot}, we model the Galactic stellar distribution using either a Sersic profile \citep{gne14} or a bulge plus disk model \citep{ken14}. 

We assume that clusters formed at redshift $z=3$, and calculate their evolution for $11.5\,$Gyr until today. As discussed in \citet{gne14}, this assumption is justified since most  globular clusters are old ($\gtrsim 10\,$Gyr) and metal-poor, and $z=3$ should be interpreted as the epoch of the peak of globular cluster formation rate. The initial mass of the clusters is drawn from a power-law distribution
\begin{equation}
  \frac{dN_{GC}}{dM_{GC}}\propto M_{GC}^{-2}, 
  \qquad M_{\rm min}< M < M_{\rm max},
\end{equation}
with $M_{\rm min}=10^4\,\mathrm{M}_{\odot}$ and $M_{\rm max}=10^7\,\mathrm{M}_{\odot}$. As shown in \citet{gne14}, the results are insensitive to the choice of $M_{\rm min}$ since low-mass clusters are expected to be quickly disrupted by the Galactic tidal field. On the other hand, the shape of the cluster mass function at present day and the total amount of cluster debris accumulated at the centre depend on the choice of $M_{\rm max}$. The parameter values above were chosen by \citet{gne14} to reproduce the observed cluster distribution.

Following their formation, globular clusters start losing mass via three mechanisms: stellar winds, dynamical ejection of stars through two-body relaxation, and stripping of stars by the Galactic tidal field. The time-dependent stellar mass loss rate has been computed by \citet{pri08} for a \citet{krp01} stellar IMF. The mass loss rate due to two-body relaxation and tidal stripping is combined as
\begin{equation}
  \frac{dM}{dt}=-\frac{M}{t_{\rm tid}}\ ,
\end{equation}
where
\begin{equation}
  t_{\rm tid}(r,M)\approx 10 \left(\frac{M}{10^5\,\mathrm{M}_{\odot}}\right)^{2/3}\, P(r)\ \mathrm{Gyr}
\end{equation}
is the typical tidal disruption time for Galactic globular clusters \citep[e.g.,][]{gie08}, where
\begin{equation}
  P(r) \equiv 41.4\left(\frac{r}{\mathrm{kpc}}\right)\left(\frac{V_c(r)}{\mathrm{km}\,\mathrm{s}^{-1}}\right)^{-1}
\end{equation}
is the normalized rotational period of the cluster orbit, which takes into account the strength of the local Galactic tidal field. Here $V_c(r)$ is the circular velocity at distance $r$ from the GC.

When a cluster arrives in the immediate vicinity of the GC, if the tidal forces are strong enough, the cluster can be torn apart since the stellar density at a characteristic place in the cluster, such as the core or half-mass radius, falls below the ambient density. Following \citet{gne14}, we adopt the average density at the half-mass radius
\begin{equation}
\rho_h=
\begin{cases}
10^3\, \mathrm{M}_{\odot}\, \mathrm{pc}^{-3}& {\rm for\;} M \le 10^5\,\mathrm{M}_{\odot} \cr
10^3 \left(\frac{M}{10^5\, \mathrm{M}_{\odot}}\right)^2\, \mathrm{M}_{\odot}\, \mathrm{pc}^{-3}& {\rm for\;} 10^5\, \mathrm{M}_{\odot} <  M < 10^6\, \mathrm{M}_{\odot} \cr
10^5\, \mathrm{M}_{\odot}\, \mathrm{pc}^{-3}& {\rm for\;} M \ge 10^6\,\mathrm{M}_{\odot}
\end{cases}
\end{equation}
The above equation limits $\rho_h$ to $10^5\,\mathrm{M}_{\odot}\,\mathrm{pc}^{-3}$ in the most massive clusters, that is about the highest observed half-mass density. A cluster is considered disrupted if the average density at the half-mass radius becomes smaller than the ambient density
\begin{equation}
  \rho_h < \rho_*(r) = \frac{V_c^2(r)}{2\pi G r^2}.
\end{equation}
Here $\rho_*(r)$ is composed both of the field stellar mass and the growing mass of the nuclear stellar cluster. As the nuclear cluster begins to build up, its stellar density will exceed even the high density of infalling globular clusters and these clusters will be directly disrupted before reaching the galaxy centre.

\citet{gne14} also included the effect of dynamical friction on cluster orbits. The orbital radius $r$ of a cluster evolves according to
\begin{equation}
  \frac{dr^2}{dt}=-\frac{r^2}{t_{\rm df}}\ ,
\end{equation}
where
\begin{equation}
  t_{\rm df}(r,M) \approx 0.23 \left(\frac{M}{10^5\ \mathrm{M}_{\odot}}\right)^{-1}\left(\frac{r}{\mathrm{kpc}}\right)^2\left(\frac{V_c(r)}{\mathrm{km}\,\mathrm{s}^{-1}}\right)\ \mathrm{Gyr}.
\end{equation}

\subsection{Galactic gravitational potential}
\label{sec:pot}

We use two different models to describe the Milky Way potential. First, we use a 4-component model "K14" \citep{ken08,ken14,fra17}
\begin{equation}
  \Phi(r)=\Phi_{BH}+\Phi_{bul}+\Phi_{disk}+\Phi_{NFW},
\end{equation}
that includes a central black hole ($M_{BH}=4 \times 10^6\,$M$_{\odot}$), a bulge ($M_{bul}=3.76\times 10^9\,$M$_{\odot}$ and $a=0.1\,$kpc; \citealt{her90}), a disk ($M_{disk}=5.36\times 10^{10}\,$M$_{\odot}$, $b=2.75\,$kpc and $c=0.3\,$kpc; \citealt{miy75}), and a dark matter halo ($M_{DM}=10^{12}\,$M$_{\odot}$ and $r_s=20\,$kpc; \citealt{nfw97}). 

We also consider a second model "G14" \citep{gne14}, where the bulge and the disk are replaced by a \citet{ser63} profile with total mass $M_{S}=5\times 10^{10}\,$M$_{\odot}$ and effective radius $R_{e}=4\,$kpc.

\begin{figure} 
\centering
\includegraphics[scale=0.55]{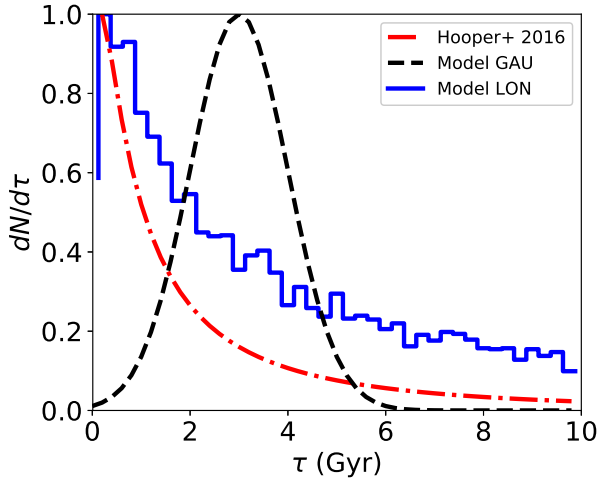}
\caption{Normalized (to the peak value) distributions of characteristic spin-down timescale for \citet{hol16} model, Model GAU-K14, and Model LON-K14. Note that the \citet{hol16} distribution peaks at $\tau\lesssim 0.5$ Gyr, in contrast with the typical values found in globular clusters, $1-5\,$Gyr \citep{fre01a,fre01b}.}
\label{fig:taus}
\end{figure}

Figure~\ref{fig1} confirms that our models match observational constraints on the mass of the Milky Way nuclear cluster and on the spatial radial distribution of the remaining globular cluster system. The left panel shows that the total mass deposited by globular clusters at the end of the simulation for both potential models is consistent with the observed mass inside $20\,$pc. The right panel shows that the number density of globular clusters surviving to the present time is consistent with the number density of observed Milky Way's globular clusters.

\subsection{Evolution of milli-second pulsars}
The gamma-ray luminosity of a given pulsar is taken to evolve with time as
\begin{equation}
L_{\gamma}(t)=\frac{L_{\gamma,0}}{[1+(t/\tau)^{1/2}]^2} ,
\label{eqn:lgamma}
\end{equation}
where $L_{\gamma,0}$ is the initial luminosity and $\tau$ is the characteristic spin-down timescale for a MSP to lose its rotational kinetic energy
\begin{equation}
\tau=\frac{E}{\dot{E}} = \frac{P}{2\dot{P}},
\label{eqn:tau}
\end{equation}
where $P$ and $\dot{P}$ are the MSP rotational period and its derivative, respectively. Recently, \citet{ole16} have argued that $L_\gamma\propto (1+(t/\tau)^{1/2})^{-1}$ is more consistent with the data, which would make MSPs slower to lose energy. The shape of the $\tau$ distribution and its relation to $L_\gamma$ turn out to be the two most important ingredients controlling the final contribution to the GC excess due to the MSPs coming from disrupted globular clusters.

\citet{fre01a, fre01b} used 10 years of observations of NGC 104 (47 Tuc) to probe the cluster's MSP population. For most of the MSPs, they were able to measure the observed period derivative $\dot{P}_{\rm obs}$. In general, $\dot{P}_{\rm obs}$ is given by the sum
\begin{equation}
 \left(\frac{\dot{P}}{P}\right)_{\rm obs} = \left(\frac{\dot{P}}{P}\right)_{\rm int} + \frac{a_1}{c} + \frac{a_2}{c} + \frac{a_3}{c},
\end{equation}
where the first term is the intrinsic MSP spin-down \citep{pra17}
\begin{equation}
c\left(\frac{\dot{P}}{P}\right)_{\rm int}=7.96\times 10^{-10}\left(\frac{B}{2\times 10^8\ \mathrm{G}}\right)^2 \left(\frac{2\ \mathrm{ms}}{P}\right)^2
\label{eqn:pdot}
\end{equation}
and the others are contamination terms. The terms $a_1$ and $a_2$ are due to the difference in Galactic acceleration between NGC 104 and the barycentre of the Solar System along the line of sight and the acceleration of the MSP in the cluster potential \citep{fre01a}, respectively, while $a_3$ is the \citet{shk70} effect due to the proper motion, which is not negligible for MSPs \citep{abd13}. After correcting for spurious contributions, \citet{fre01a,fre01b} found that the average characteristic age of the pulsars is larger than $\approx 1$ Gyr, and no MSP was detected with $\tau \lesssim 200$ Myr.

\citet{hol16} estimated that, if the spin-down was applied to the MSPs deposited by globular clusters, such a population would account only for a few percent of the total GC excess. However, their calculation contains two caveats. First, their model did not consider the history of globular cluster formation and disruption in the GC. Clusters inspiral in the GC at different times depending on their mass, and where they are formed in the Galaxy. Second, they draw $\tau\propto L_\gamma^{-1}$ and, more importantly, their distribution of spin-down times is inconsistent with observational constraints on the $\tau$ distribution in globular clusters \citep{fre01a}. Figure~\ref{fig:taus} shows that the distribution of $\tau$ (normalized to the peak value) used in \citet{hol16} peaks at $\tau\ll 1$ Gyr, in contrast with the observations of \citet{fre01a}. Thus, by adopting the small values of $\tau$, the \citet{hol16} model overestimates the effect of spin-down and underestimates the total gamma-ray luminosity in the GC.

Figure~\ref{fig:taus} also shows two additional distributions of $\tau$ (normalized to the peak value), which we use for our models. In model GAU, we adopt a Gaussian distribution with mean of $3\,$Gyr, consistent with \citet{fre01a}, who found a characteristic age of $\approx 3\,$Gyr for MSPs in NGC 104. In model LON, we compute the spin-down rate via Eq.~\ref{eqn:pdot}, by sampling magnetic fields from a log-normal distribution with mean of $10^{8.47}\,$G and standard deviation of $0.33$ and by sampling periods from a log-normal distribution with $P_0 = 3\,$ms and $\sigma_P = 0.234$ \citep{pra17}. In this model, the $\tau$ distribution has a mean around $1$~Gyr, but also a non-negligible tail at larger $\tau$. Since our distributions predicts larger values of $\tau$ than in \citet{hol16}, they produce lower spin-down rates and higher total gamma-ray luminosity.

Table~\ref{tab2} lists all the models considered in this work. For each model GAU and LON, we run two variants with the \citet{ken14} and \citet{gne14} potentials. For comparison, we also run a model with no MSP spin-down (Model NSD).

\begin{table*}
\caption{Models: name, Galactic potential (GP), spin-down ($\tau$), slope of the luminosity distribution ($\alpha$), gamma-ray luminosity-to-mass ratio ($L_{\gamma}/M_{GC}$).}
\centering
\begin{tabular}{lcccc}
\hline\\[-2mm]
Name & GP & $\tau$ (Gyr) & $\alpha$ & $L_{\gamma}/M_{GC}$\\[1mm]
\hline\\[-2mm]
Model NSD		& \citet{ken14}	& -													& $0.5$-$1$ 	& Eq. \ref{eqn:lgmcl}\\
Model GAU-K14	& \citet{ken14}	& \citet{fre01a,fre01b}								& $0.5$-$1$	& Eq. \ref{eqn:lgmcl}\\
Model GAU-G14	& \citet{gne14}	& \citet{fre01a,fre01b}								& $0.5$-$1$ & Eq. \ref{eqn:lgmcl}\\
Model GAU-K14C	& \citet{ken14}	& \citet{fre01a,fre01b}								& $0.5$-$1$	& const\\
Model LON-K14	& \citet{ken14}	& \citet{pra17}+Eqs.~\ref{eqn:pdot}	& $0.5$-$1$	& Eq. \ref{eqn:lgmcl}\\
Model LON-G14	& \citet{gne14}	& \citet{pra17}+Eqs.~\ref{eqn:pdot}	& $0.5$-$1$ & Eq. \ref{eqn:lgmcl}\\
Model LON-K14C	& \citet{ken14}	& \citet{pra17}+Eqs.~\ref{eqn:pdot}	& $0.5$-$1$	& const\\
\hline
\end{tabular}
\label{tab2}
\end{table*}

\begin{table*}
\caption{Globular Cluster: name, mass (M), gamma-ray flux (F), distance (D), power index ($\alpha$) and cutoff energy ($E_{\rm cut}$).}
\centering
\begin{tabular}{lccccc}
\hline\\[-2mm]
Name & M ($\mathrm{M}_{\odot}$) & F (erg cm$^{-2}$ s$^{-1}$) & D (kpc) & $\alpha$ & $E_{\rm cut}$\\[1mm]
\hline\\[-2mm]
NGC 104				& $7.12\times 10^5$ & $2.436\times 10^{-11}$ & 4.5 & 1.18 &2.51\\
NGC	2808			& $6.93\times 10^5$ & $3.546\times 10^{-12}$ & 9.6 & 1.36 &3.16\\
NGC	5139			& $1.54\times 10^6$ & $5.900\times 10^{-12}$ & 5.2 & -0.12 &1.26\\
NGC	5904			& $4.06\times 10^5$ & $2.131\times 10^{-12}$ & 7.5 & 1.86 &3.98\\
NGC	6093			& $2.38\times 10^5$ & $3.986\times 10^{-12}$ & 10.0 & 1.38 &5.01\\
NGC	6139			& $2.68\times 10^5$ & $5.330\times 10^{-12}$ & 10.1 & 2.28 &19.95\\
NGC	6218			& $1.02\times 10^5$ & $2.969\times 10^{-12}$ & 4.8 & 2.24 &$\geq$ 100\\
NGC	6266			& $5.70\times 10^5$ & $1.710\times 10^{-11}$ & 6.8 & 1.36 &3.16\\
NGC	6316			& $2.63\times 10^5$	& $1.091\times 10^{-11}$ & 10.5 & 2.00 &7.94\\
NGC	6342			& $4.49\times 10^4$ & $4.339\times 10^{-12}$ & 8.5 & 2.16 &15.85\\
NGC	6388			& $7.05\times 10^5$ & $1.732\times 10^{-11}$ & 9.9 & 1.52 &3.16\\
NGC	6397			& $5.51\times 10^4$ & $6.390\times 10^{-12}$ & 2.3 & 2.90 &50.12\\
NGC	6440			& $3.84\times 10^5$	& $2.392\times 10^{-11}$ & 8.5 & 2.32 &10.00\\
NGC	6441			& $8.65\times 10^5$ & $1.252\times 10^{-11}$ & 11.6 & 2.04 &10.00\\
NGC	6541			& $3.11\times 10^5$ & $3.251\times 10^{-12}$ & 7.5 & 1.16 &2.51\\
NGC	6652			& $5.60\times 10^4$	& $4.495\times 10^{-12}$ & 10.0 & 1.38 &3.16\\
NGC	6717			& $2.23\times 10^4$	& $1.816\times 10^{-12}$ & 7.1 & 0.38 &2.51\\
NGC	6752			& $1.50\times 10^5$ & $2.866\times 10^{-12}$ & 4.0 & 0.12 &0.79\\
NGC	7078			& $5.75\times 10^5$ & $3.160\times 10^{-12}$ & 10.4 & 2.42 &6.31\\
Palomar 6			& $6.30\times 10^4$ & $5.489\times 10^{-12}$ & 5.8 & 0.94 &1.26\\
Terzan 5			& $1.13\times 10^5$ & $5.973\times 10^{-11}$ & 6.0 & 1.16 &2.51\\
2MASS-GC01			& $3.38\times 10^4$ & $2.476\times 10^{-11}$ & 3.1 & 1.06 &1.26\\
2MASS-GC02			& $1.07\times 10^4$ & $8.846\times 10^{-12}$ & 3.9 & 1.08 &1.26\\
GLIMPSE C02			& -					& $1.630\times 10^{-11}$ & 4.6 & 1.94 &7.94\\	
GLIMPSE C01			& $2.81\times 10^4$ & $9.020\times 10^{-12}$ & 5.0 & -0.74 &1.58\\			
\hline
\end{tabular}
\label{tab1}
\end{table*}

\subsection{MSPs from globular cluster debris}

\begin{figure} 
\centering
\includegraphics[scale=0.55]{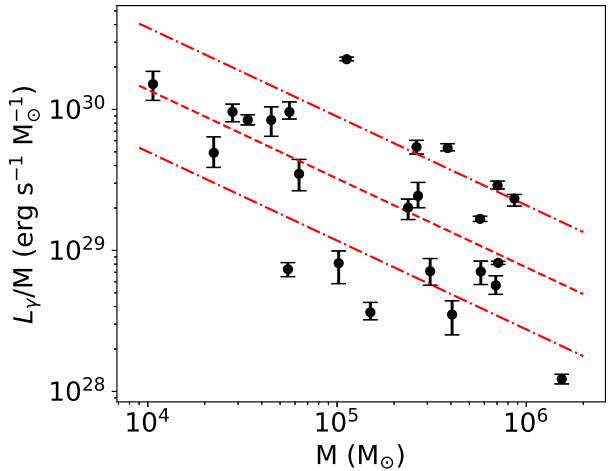}
\caption{Ratio of gamma-ray luminosity to globular cluster mass $L_{\gamma}/M_{GC}$ as a function of cluster mass. Dashed line shows the best log-linear fit, dot-dashed lines show the $1\,\sigma$ deviations.}
\label{fig:lgammass}
\end{figure}

Table~\ref{tab1} lists the mass and gamma-ray flux of Galactic globular clusters, according to the analysis of \textit{Fermi} data by \citet{hol16}. To calculate the globular cluster masses, we convert the absolute V magnitudes of \citet{har96} (2010 edition) by assuming a typical mass-to-light ratio of $1.5\,\mathrm{M}_{\odot}/\mathrm{L}_{\odot}$ \citep{har17}. No absolute V magnitude was found for GLIMPSE C02. Distances for 2MASS-GC01 and 2MASS-GC02 are taken from \citet{bic08}, for GLIMPSE C02 and GLIMPSE C01 from \citet{kur08} and \citet{dav11}, respectively, and the rest are from \citet{hol16}. Table~\ref{tab1} also gives the best fit parameters of the cluster energy spectra of the form $dN_\gamma/dE_\gamma\propto E_\gamma^{-\alpha}\exp(-E_\gamma/E_{\rm cut})$ \citep{hol16}.

From the data in Table~\ref{tab1}, we calculate the ratio between the gamma-ray luminosity $L_{\gamma}$ and mass $M_{GC}$ for each cluster. Figure~\ref{fig:lgammass} shows the ratio $L_{\gamma}/M_{GC}$ as a function of cluster mass. We fit the data with the log-linear relation
\begin{equation}
  \log (L_{\gamma}/M_{GC})=32.66\pm 0.06-(0.63\pm 0.11)\log M_{GC}.
\label{eqn:lgmcl}
\end{equation}
We note that Palomar 6, GLIMPSE C01, and 2MASS-GC02 have relative large errors ($\approx 25\%, 13\%, \mathrm{and}\, 25\%$, respectively) in the determination of the gamma-ray flux (see \citealt{hol16}). We repeated the fit without these three clusters and found that the resulting log-linear relation does not change significantly.

\citet{bra15} used the $2\,$GeV fluxes of 11 globular clusters as measured by \citet{cho15} to estimate an average flux density per unit mass at $8.3\,$kpc. They used this average flux to scale the present-day stellar mass deposited by clusters in the GC as calculated in \citet{gne14}. The calculation of \citet{bra15} did not take into account that $L_{\gamma}\propto M^{0.37}$ according to Eq.~\ref{eqn:lgmcl}, and that after a MSP is deposited in the GC its gamma-ray luminosity decreases in time according to Eq.~\ref{eqn:lgamma}. In our models, we follow the gradual disruption of globular clusters across cosmic time and use Eq.~\ref{eqn:lgmcl} to infer the MSP population left by a cluster of given mass. As MSPs are deposited in the GC, we evolve their initial luminosity taking into account energy loss due to magnetic-dipole breaking, according to Eq.~\ref{eqn:tau}. This allows us to calculate and compare the present day gamma-ray luminosity with the \textit{Fermi} GC excess data.

\section{Results}

We evolve the globular cluster population according to the model in \S\ref{sec:model} and compute the mass deposited by each globular cluster as a function of time $t$ and radius $r$ from the GC. At each time $t$  we compute the amount of gamma-ray luminosity expected from all MSPs left in the cluster debris, $L_{\gamma,\rm tot}^{\rm dep}(t)$, using the mass-to-light ratio distribution of Eq.~\ref{eqn:lgmcl}. We then generate a sample of individual MSPs by drawing from the power-law distribution
\begin{equation}
  \frac{dN}{dL_{\gamma}}\propto L_{\gamma}^{-\alpha}
  \label{eqn:lgdist}
\end{equation}
between $L_{\gamma,\rm min}=10^{31}\,\mathrm{erg\, s}^{-1}$ and $L_{\gamma,\rm max}=10^{36}\,\mathrm{erg\, s}^{-1}$, which are approximately the minimum and maximum observed MSP luminosities \citep{aje17}. We sample from the above distribution until the total luminosity from the deposited MSPs equals $L_{\gamma,\rm tot}^{\rm dep}(t)$. This gives us the number of MSPs, $N_{\rm MSP}(t)$. Note that the slope and limits of the MSP luminosity distribution are uncertain, mainly because of the incompleteness of the pulsar sample and variations in detection efficiency across the sky \citep{pet15,bar16}. As a reference, \citet{aje17} found $\alpha=1.20\pm 0.08$ by fitting the field pulsar data \citep{man05,abd13}.

We assume that the formation and disruption rates of MSPs while in bound clusters balance each other, so that the number of MSPs per unit luminosity is constant until the cluster is fully disrupted by the Galactic tidal field. After a MSP is released from its parent cluster we evolve its spin period and luminosity according to the value of $\tau$ drawn from the adopted distribution. Specifically, we evolve the initial luminosity of each MSP by means of Eq.~\ref{eqn:lgamma}, respectively, from the time $t$ when it is deposited in the GC up to the present time of $\approx 13.7$ Gyr. Finally, we sum over the luminosity of all deposited MSPs and compare the spatial and luminosity profiles with the observations. To improve the statistical significance of the results, we average over $1000$ random realizations of the model. We assume a GC distance of $8.3$~kpc. 

\subsection{The predicted gamma-ray flux}

We first describe the results for the NSD, GAU-K14, and LONG-K14 models with $\alpha=1$, and then consider variations due to different choices of the model parameters $\alpha$, $M_{\rm max}$, and $L_{\gamma,\rm max}$.

\begin{figure*}
\centering
\begin{minipage}{20.5cm}
\subfloat{\includegraphics[scale=0.55]{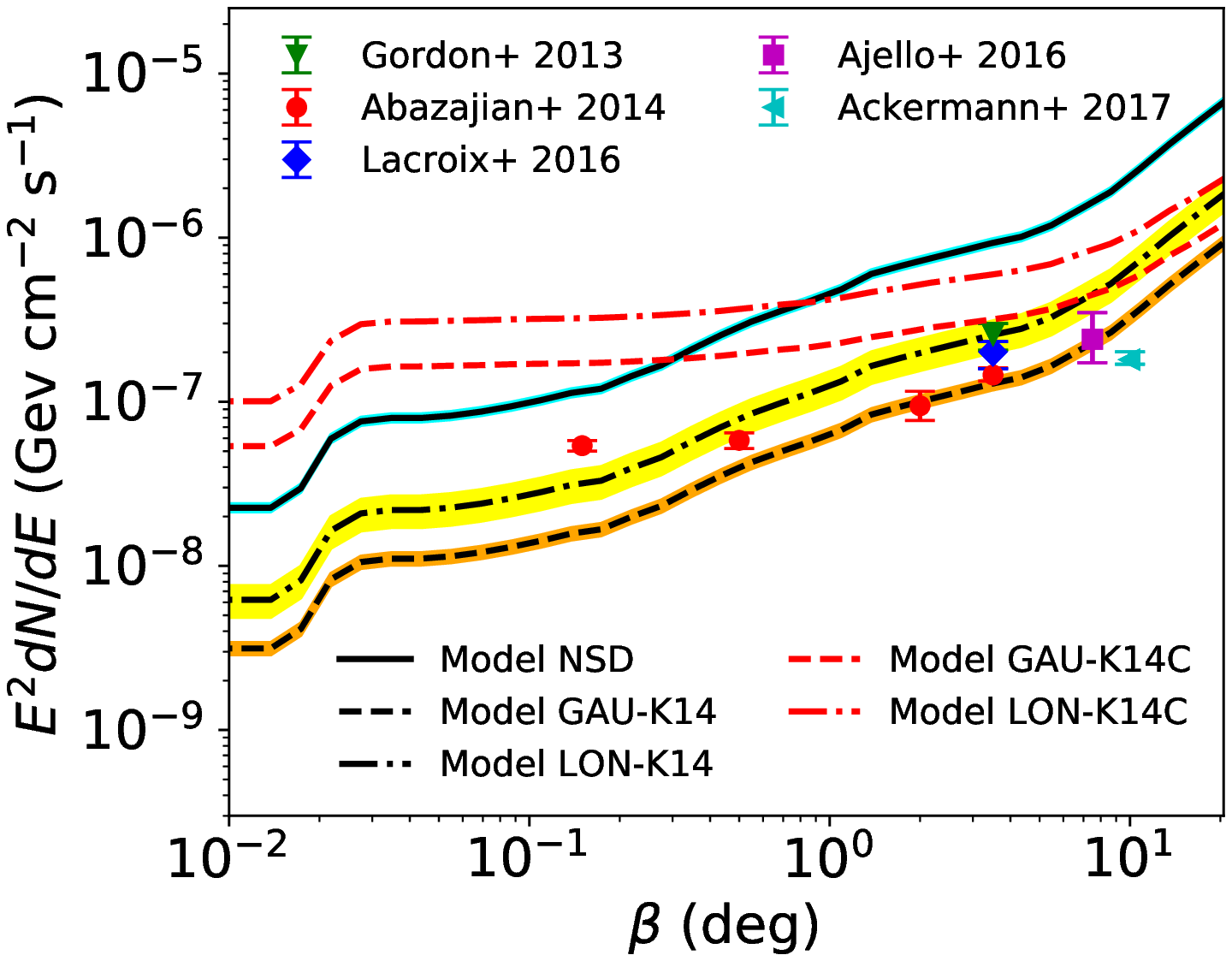}}
\subfloat{\includegraphics[scale=0.55]{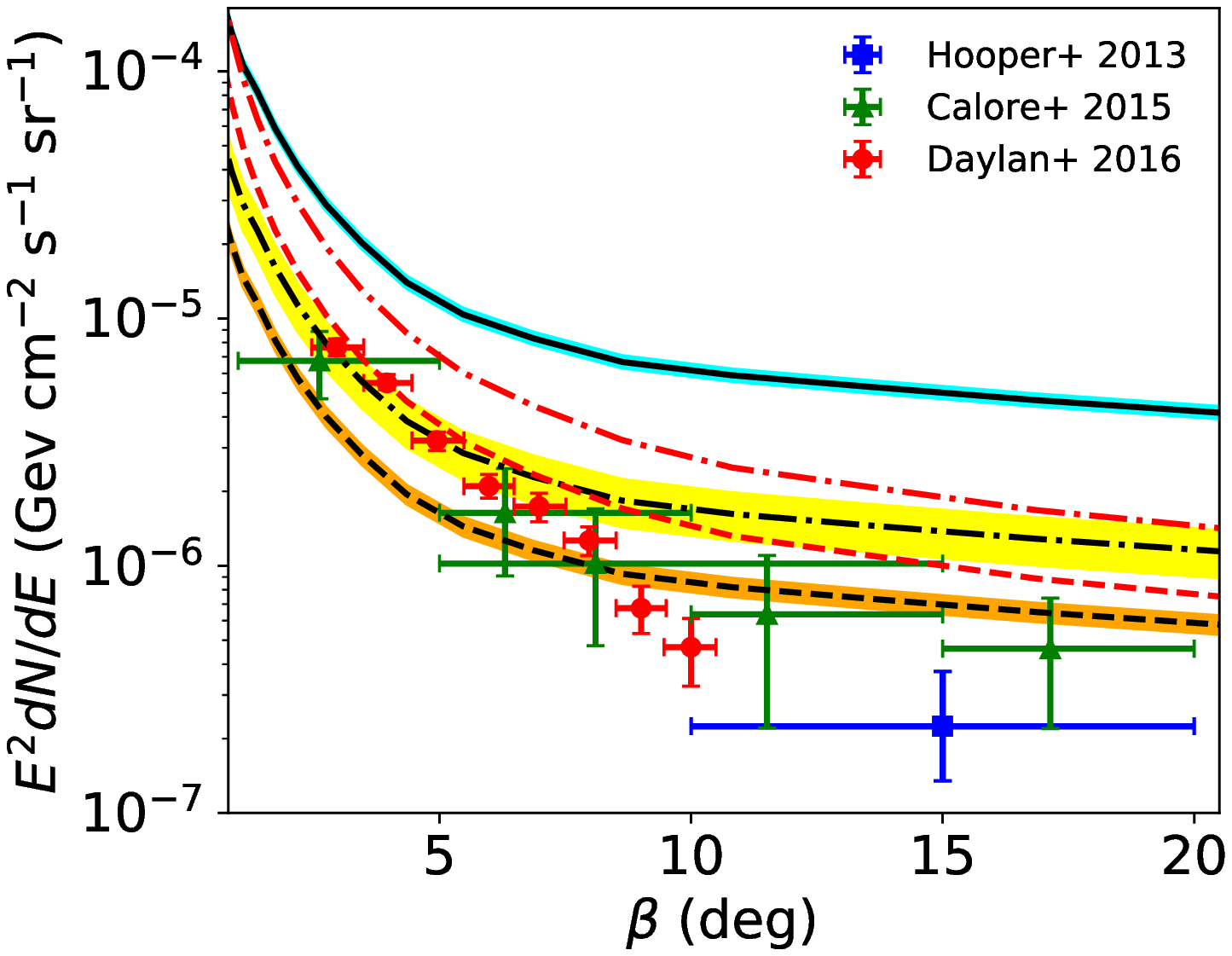}}
\end{minipage}
\caption{Predicted MSP integrated gamma-ray flux within angle $\beta$ from the GC (left panel) and  differential flux at angle $\beta$ (right panel) at $2$~GeV, in the case $\alpha=1$. Shaded regions show the $3\sigma$ deviations. Symbols show results of the published analysis of the \textit{Fermi} data by \citet{gor13, hos13, aba14, cal15, aje16, day16, lac16, ack17}.}
  \label{fig:fluxg}
\end{figure*}

The gamma-ray luminosity of a MSP decreases (Eq.~\ref{eqn:lgamma}) as a consequence of the energy loss due to magnetic-dipole emission as $L_\gamma\propto (1+(t/\tau)^{1/2})^{-2}$. Consequently, a period distribution peaked at small values and without a considerable tail at high $\tau$'s, as in \citet{hol16}, implies that each MSP spins down very rapidly and has a final luminosity that is only a few percent of its initial value. Our Model GAU  is a Gaussian peaked at $\tau=3$~Gyr, while Model LON distribution is peaked at $\tau\approx 1$~Gyr. Hence, on average, MSPs deposited in the GC more than a few Gyr ago will have had enough time to spin-down considerably and would not contribute much to the total final luminosity. However, in both models there is a non-negligible tail at $\tau\gtrsim 5\,$Gyr so that the final luminosity will be dominated by ancient MSPs with long spin-down timescales.

In Fig.~\ref{fig:fluxg}, we compare our predicted integrated and differential gamma-ray flux at $2$~GeV with the analysis of the \textit{Fermi} excess data reported in the literature \citep{gor13, hos13, aba14, cal15, aje16, day16, lac16, ack17}. Different authors adopt different methods for data analysis and focus on different regions of interest (ROI), i.e. the width of the region centered on the GC in which they collect data. Some authors present integrated flux analysis in the ROI, while others prefer differential fluxes.

Fig.~\ref{fig:fluxg} shows that the gamma-ray flux predicted by Model NSD, in which the spin-down is not taken into account, is about one order of magnitude larger than in other models and in observations. Therefore, Model NSD is ruled out by current data. On the other hand, Models GAU-K14 and LONG-K14 are consistent with the data throughout the whole relevant radial range, except at very small angles ($\beta\approx 0.2^\circ$) where our model luminosity is about a factor of two smaller than observed. Note that the complex models of interstellar emission subtracted from the raw \textit{Fermi} data produce large systematic uncertainty within the ROI \citep{hos13,aba14,cal15}. Errors arise from the assumptions used to derive the model and may potentially have a large effect on the characterization of the different gamma-ray sources in \textit{Fermi} data \citep{aba14}. The Galactic diffuse background is the dominant component and confusion source within the ROI \citep{zho15,hua16}. Moreover, modeling of the bremsstrahlung emission of high-energy electrons interacting with the molecular gas contributes additional uncertainty. When the GC excess is interpreted with MSP emission, \citet{pet15} showed that another relevant source of uncertainty may be the endpoint of the MSP luminosity function. \citet{cal15} found that the spatial profile of the excess is well described by a power-law $r^{-\zeta}$, with $\zeta\approx 2.2-2.9$, and \citet{cal16} pointed out that variations of the spatial index $\zeta$ within the $1\sigma$ range could affect the total gamma-ray luminosity by up to 40\%. Given the uncertainties in data analysis and modeling, we conclude that the radial profile and normalization of the gamma-ray flux predicted by Models GAU-K14 and LONG-K14 are in reasonable good agreement with current observational constraints.

\subsection{Model uncertainties}

In order to explore the effect of other uncertain factors and model parameters, we run additional simulations varying the Galactic potential, the extremes of the MSP gamma-ray luminosity, and the extremes of the globular cluster mass function (see Tab.~\ref{tab2}).

The models in which we consider the G14 potential model or a different minimum gamma-ray luminosity $L_{\gamma,\rm min}$ do not lead to substantial differences with respect to the results presented in Fig.~\ref{fig:fluxg}. \citet{bar16} found that a power-law luminosity function with $\alpha=1.5$ and $L_{max}=7\times 10^{34}$ describes the photon clustering in the data due to the the GC excess. On the other hand, \citet{pet15} argued that a $50$\% change in maximum luminosity $L_{\gamma,\rm max}$ can translate into about one order of magnitude change in the expected number of detected sources. We repeated our calculations with $L_{\gamma,\rm max}$ reduced to $10^{35}\, \mathrm{erg\, s}^{-1}$, and found that even in this case the total gamma-ray luminosity decreases at most by a factor of two. The decrease can be readily explained by noting that MSPs with large luminosity dominate the total gamma-ray emission, which are less abundant if $L_{\gamma,\rm max}=10^{35}\, \mathrm{erg\, s}^{-1}$.

We also explored the effect of the maximum globular cluster mass. We considered the range $M_{\rm max}= 5\times 10^6 - 2\times 10^7\, \mathrm{M}_{\odot}$ \citep[as in][]{gne14} and found that the total gamma-ray luminosity increases at most by about a factor of two when $M_{\rm max}=2\times 10^7\, \mathrm{M}_{\odot}$. Our results are insensitive to the choice of $M_{\rm min}$, since light clusters are expected to be rapidly disrupted by the Galactic tidal field before reaching the GC \citep{gne14}.
In order to quantify how sensitive our results are to the assumed fraction of the nuclear cluster mass coming from globular clusters, we also evolved a  model with $f_{GC,i}=0.008$. With this choice, the total mass transported in the inner 20 pc was half that in our fiducial models, while the total predicted gamma-ray emission (within 10 deg) decreased only by 3\%.

We note that Eq. \ref{eqn:lgmcl} predicts $L_{\gamma}\propto M_{GC}^{0.37}$. However, the correlation between $L_{\gamma}$ and the present-day cluster mass may not hold for the initial population. The evolution of the MSP population of globular clusters, and hence of their gamma-ray luminosity, is still not well understood and deserves future work. To test the dependence of the adopted $L_{\gamma}-M_{GC}$ relation on our results, we run our models with $L_\gamma/M_{GC} = \mathrm{const}\, = 4.57\times 10^{29}$~erg~s$^{-1} M_\odot^{-1}$ (from Table~\ref{tab1}). Fig.~\ref{fig:fluxg} illustrates that both model GAU-K14C and model LON-K14C tend to overestimates the GC gamma-ray excess by a factor of $\sim 2$ and $\sim 5$, respectively.

{\color{red}
%Recently, \citet{fel17} studied the metallicity of stars in the inner Galaxy and concluded that their data can discard a scenario in which the majority of  
%the stellar mass in the inner parsec  
%was brought in by infalling globular clusters. 
%However, the metallicity and ages of stars over 
%larger spatial scales remain essentially unconstrained by
%current data.
%However, \citet{pfu11} showed that the majority ($\approx 80$ \%) of stars formed more than $\approx 5$ Gyr ago. 
%\citet{ant12} performed an analysis of the K-band luminosity of the stars within $\approx 30$ pc of the GC and showed that a model in which $\approx 50$ \% of the stars come from infalling globular clusters is consistent with current observational constraints.
}

\begin{figure} 
\centering
\includegraphics[scale=0.55]{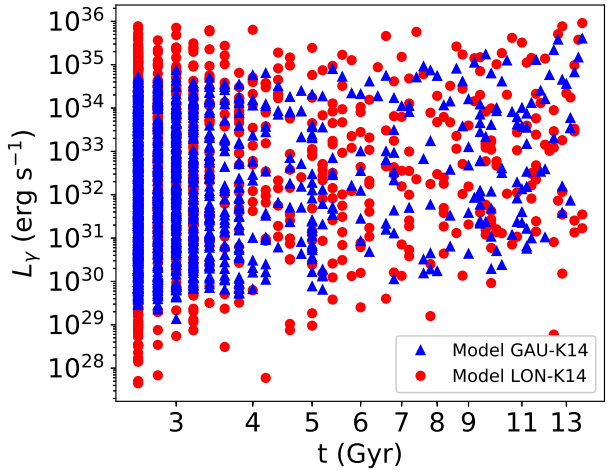}
\caption{Final $L_{\gamma}$ of MSPs left by globular clusters within $\approx 20^\circ$ as a function of the time they were left in the GC for Model GAU-K14 and Model LON-K14 for $\alpha=1$.}
\label{fig:lgammatau}
\end{figure}

\begin{figure}
\centering
\includegraphics[scale=0.55]{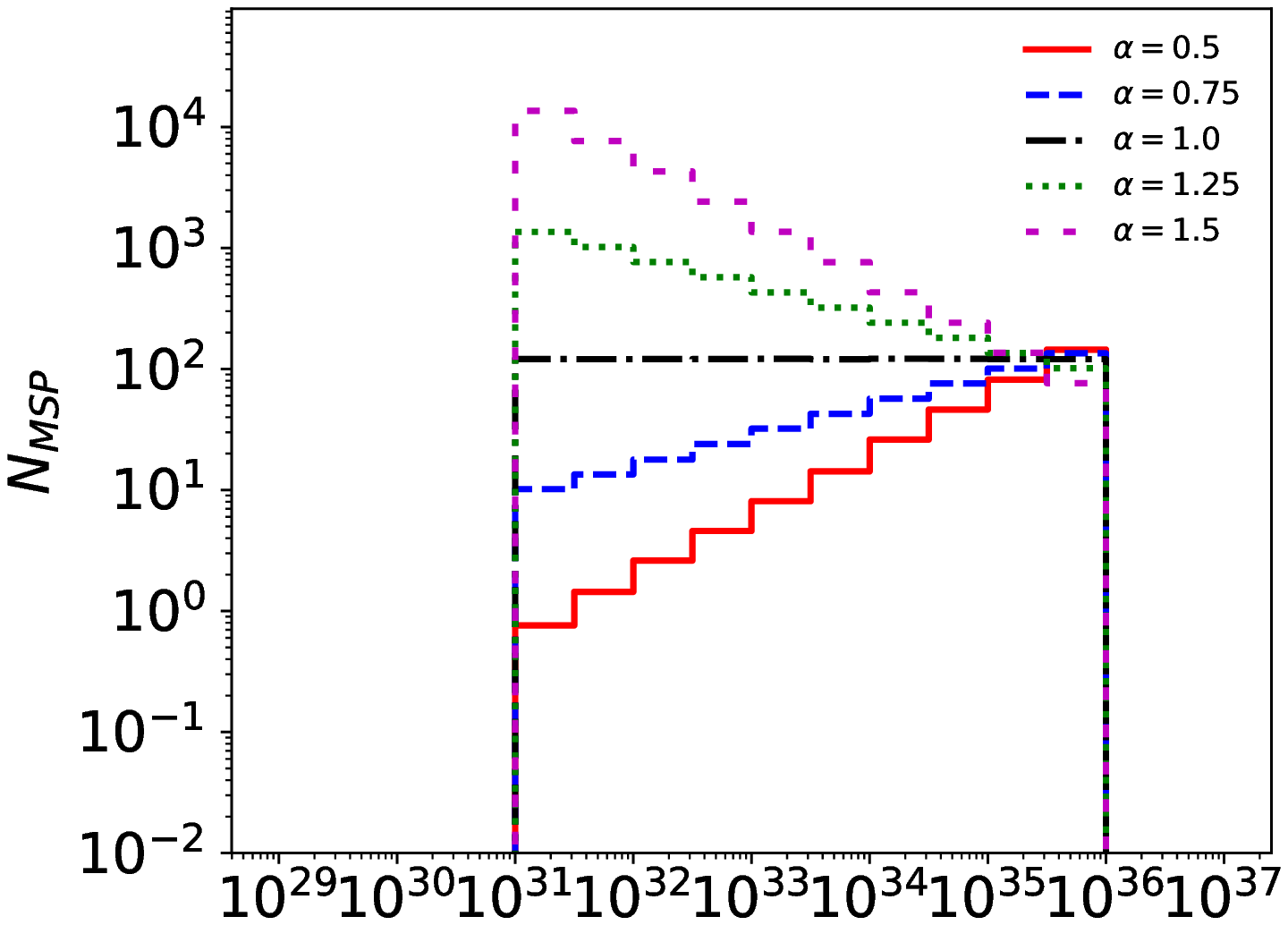}
\includegraphics[scale=0.55]{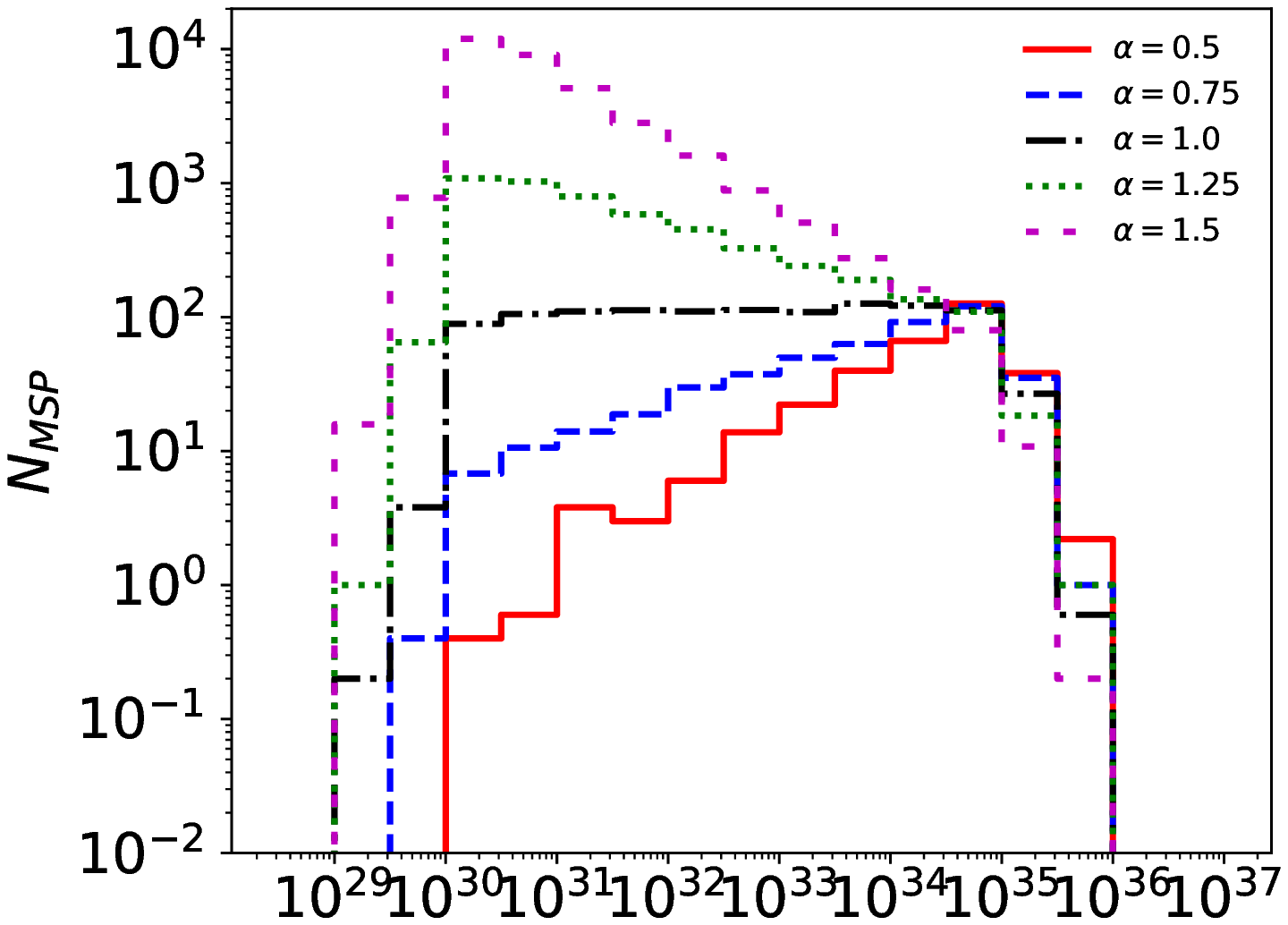}
\includegraphics[scale=0.55]{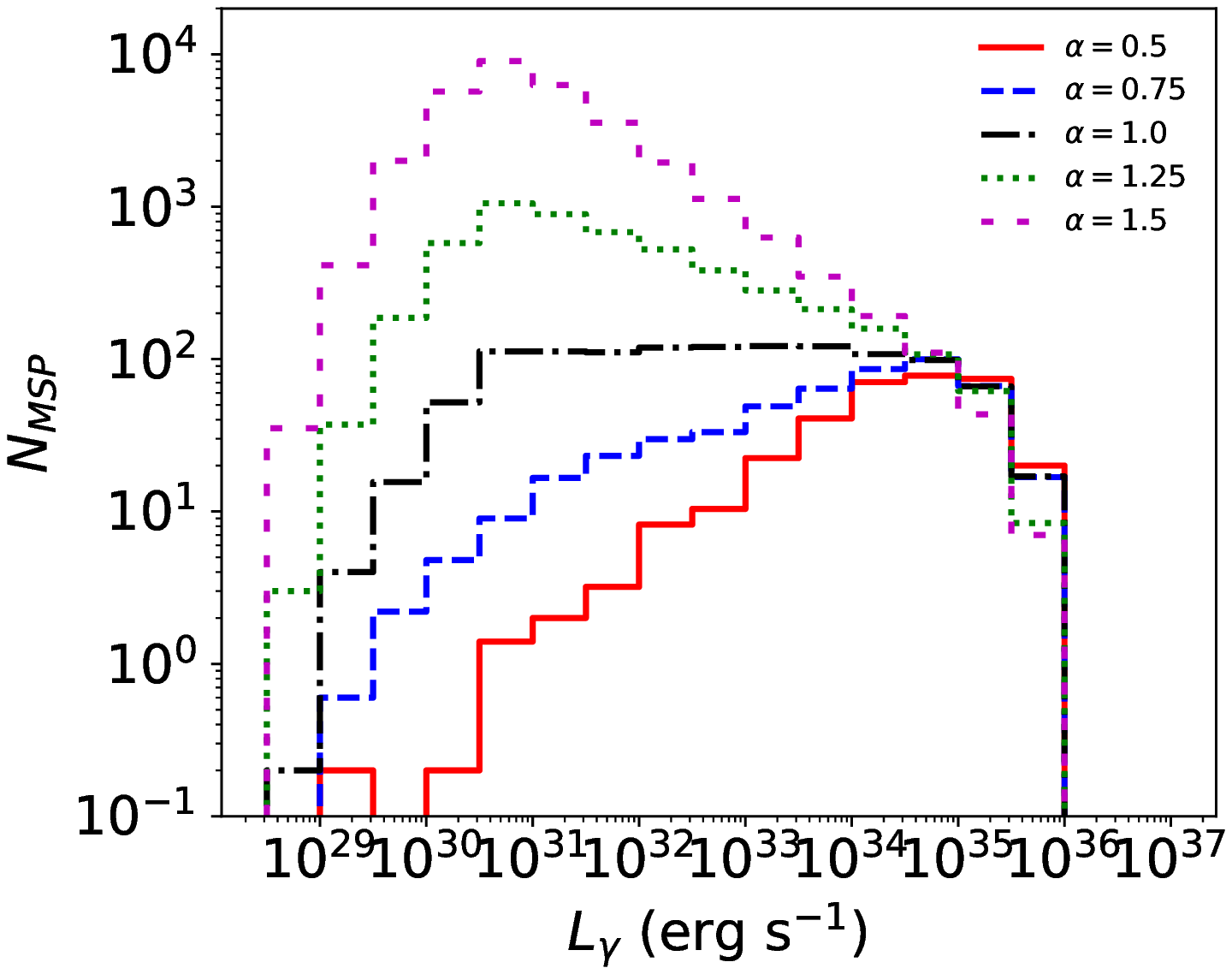}
\caption{Final distributions of MSP $L_\gamma$ for Models NSD (top), GAU-K14 (centre), and LON-K14 (bottom). Lines correspond to different values of slope $\alpha$ of the MSP gamma-ray luminosity distribution (Eq.~\ref{eqn:lgdist}).}
\label{fig:nmsp}
\end{figure}

\begin{figure}
\centering
\includegraphics[scale=0.55]{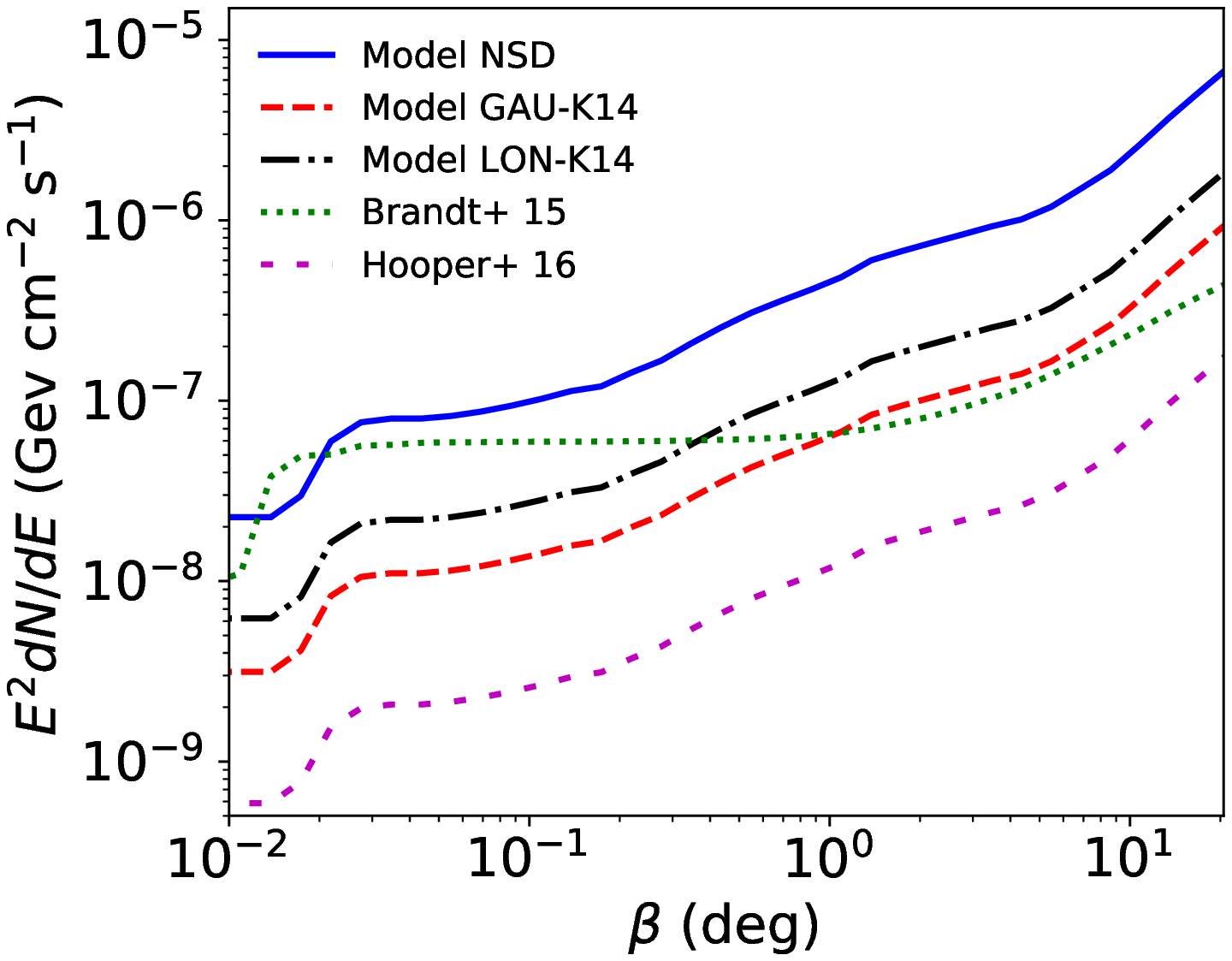}
\caption{Integrated flux within angle $\beta$ from the GC for our models as well as for \citet{bra15} model and \citet{hol16} model with the assumed $\tau$ distribution from Fig.~\ref{fig:taus}.}
\label{fig:fluxcomp}
\end{figure}

As discussed above, a key ingredient in the determination of the final gamma-ray luminosity is the characteristic spin-down $\tau$. Figure \ref{fig:lgammatau} shows the final $L_\gamma$ of MSPs as function of the time they were left in the GC by infalling clusters for Model GAU-K14 and Model LON-K14 for $\alpha=1$. In both models, about half of the total final  luminosity comes from MSPs that were born in the first few Gyr. Moreover, while Model GAU-K14 predicts that the most luminous MSPs were left up to $\approx 3$~Gyr ago, Model LON-K14 also contains MSPs that were left in the GC in the first Gyr but remained very luminous because of the long tail of $\tau$. Moreover, the relation between $L_\gamma$ and $\tau$ via Eq.~\ref{eqn:lgamma} is also important. Note that we assume the spin-down timescale to be independent on $L_\gamma$, while \citet{hol16} assumed $\tau\propto L_\gamma^{-1}$. Recently, \citet{ole16} have argued that $L_\gamma\propto (1+(t/\tau)^{1/2})^{-1}$ is more consistent with that data, which makes MSPs less efficient in losing energy and gives a larger GC gamma-ray excess.

Figure~\ref{fig:nmsp} illustrates the final luminosity distribution of MSPs for our three models with the K14 Galactic potential. In all models with different $\tau$ distribution, varying the power-law slope $\alpha$ leads to modification of the shape (flattening) of the luminosity distribution at small $L_{\gamma}$. However, the high-luminosity end ($L_{\gamma}\gtrsim 10^{34}\, \mathrm{erg\, s}^{-1}$) is not significantly affected. Since the less luminous MSPs do not contribute much to the gamma-ray excess, which is dominated by the few most luminous MSPs, our conclusions are insensitive to the particular choice of $\alpha$.

\section{Discussion}

The predicted gamma-ray flux and spatial distribution of MSPs from globular cluster disruption are both consistent with the GC excess data. Figure~\ref{fig:fluxcomp} compares the integrated flux within angle $\beta$ from the GC for our three fiducial  models (with $\alpha=1$) to the \citet{bra15} and \citet{hol16} results. 

\citet{bra15} computed the gamma-ray luminosity per unit stellar mass for the globular clusters using the $2$ GeV flux data measured by \citet{cho15}. They assumed the same gamma-ray luminosity per unit mass as for intact clusters (average $2$ GeV flux density at $8.3$~kpc of $2\times 10^{-15}$~GeV cm$^{-2}$ s$^{-1}$ $\mathrm{M}_{\odot}^{-1}$) and predicted the integrated flux in $\gamma$-rays without additional parameters by scaling the total mass deposited by clusters in the GC with the average gamma-ray luminosity. \citet{hom16} argued that the model of \citet{bra15} did not take into account the spin-down of MSPs, and  that if this effect was included in the calculation only of a few percent of the GC gamma-ray excess may be explained. As noted by \citet{hol16}, when a globular cluster is disrupted and its MSP population deposited in the GC, this population evolves losing their rotational energy through magnetic dipole breaking, becoming less luminous with time.

Unlike \citet{bra15}, we did consider the history of the MSP population deposited into the GC as their parent globular clusters lose mass and finally dissolve across the Milky Way's history. We have taken into account that the gamma-ray luminosity contributed by each cluster depends on its mass through the mass-to-light distribution Eq.~\ref{eqn:lgmcl}. These new ingredients preclude a direct comparison of our Model NSD to the results of \citet{bra15}, and explain why the two models do not lead to similar gamma-ray amplitude and spatial distribution. In addition to this, we evolved the deposited MSP population by taking into account their energy loss in our Model GAU and Model LON. Our study shows that taking into account both the 
details of the merger hostory of the globular clusters and the MSP spin-down is critical  to determining the final shape of the gamma-ray excess as due to MSPs.

On the other hand, the \citet{hol16} model assumed a spatial profile $dN/dV\propto r^{-2.4}$ for MSPs and used $\tau\propto L_\gamma^{-1}$ and a distribution of spin-down timescales (see Fig.~\ref{fig:taus}) peaked at $\approx 0.25$~Gyr. While the spatial template is roughly consistent with the deposited mass by globular clusters, the spin-down distribution is in tension with observations \citep{fre01a,fre01b}. Consequently, the \citet{hol16} model underestimates the final MSP gamma-ray luminosity by roughly one order of magnitude.

Regarding the observability of individual MSPs responsible for the GC excess, \citet{cho15} claimed that nearly 60 Galactic bulge pulsars should already have been present in the \textit{Fermi-LAT} catalog, if the excess is due to MSPs. \citet{lee15} and \citet{bar16} published evidence for the existence of an unresolved population of $\gamma$-ray sources in the inner $\approx 20^\circ$ of the Galaxy, with a total flux and spatial distribution consistent with the \textit{Fermi} excess. These faint sources have been interpreted as belonging to the Galactic bulge pulsar population. MSPs could also be observed in radio frequencies. \citet{cal16} investigated the possibility of detection of this pulsar population in radio waves, but found that current radio telescopes are not sensitive enough to radio flux of MSPs in the bulge.

Recently, \citet{aje17} studied the sensitivity of the \textit{Fermi}-LAT for detecting pulsars as point sources in the GC. They concluded that the \textit{Fermi} detection efficiency is low near the Galactic plane because of the bright foregrounds, while $\gamma$-ray pulsars within a few degrees of the GC are hard to detect due to source confusion. For source densities $\lesssim 5$ deg$^{-2}$, the efficiency is around 50\%, and lower at higher density. Their maximum likelihood analysis suggested that a population of $\approx 2.7$ gamma-ray pulsars in the Galactic disk for each pulsar in the Galactic bulge (distributed as $dN/dV\propto r^{-2.6}$) is consistent with the number of known pulsars and related sources in the \textit{Fermi} catalog. This model is preferred at the level of $7$ standard deviations with respect to a model with MSPs only in the disk. They concluded that, if MSP emission have to explain the GC excess, the Galactic bulge pulsar population must include $\approx 500- 2300$ sources, most of them unresolved. Moreover, they found that the dark matter interpretation of the gamma-ray excess is highly disfavored since it is not consistent with the observed distribution of pulsar candidates.

\section{Conclusions}

The \textit{Fermi-LAT} telescope provided evidence of the gamma-ray excess toward the region surrounding the GC, peaked at $\approx 2$ GeV. The excess has a spherical density distribution, similar to that of a typical dark matter halo. This similarity has led to the interpretation of the signal as being due to dark matter particle annihilation. An alternative explanation to the excess is that it comes from the emission of thousands of unresolved MSPs.

In this paper, we discussed in detail the possibility that the \textit{Fermi} GC excess is due to an unresolved population of MSPs in the bulge. Such MSPs come from the debris of globular clusters that were disrupted by the Milky Way's tidal field, as suggested by \citet{bra15}. We used a semi-analytical model to calculate the formation, migration, and disruption of globular clusters throughout the age of the Galaxy. We considered several variants of the model and showed that the most important ingredient in this scenario is the typical MSP spin-down rate. If it is not included, the resulting total $\gamma$-ray luminosity predicted by the globular cluster model is about one order of magnitude larger than the observed \textit{Fermi} excess. When we take into account the pulsar spin-down and use a distribution of characteristic periods consistent with that of observed MSPs in globular clusters, the model reproduces the observed radial distribution and total luminosity of the excess, giving a natural astrophysical motivation to the GC gamma-ray excess.

\section{Acknowledgements}

We acknowledge Timothy Brandt, Tim Linden, Re'em Sari, Roberto Capuzzo-Dolcetta, Manuel Arca-Sedda for useful discussions, and anonymous referee for constructive comments. GF acknowledges Moharana Reetanjali for useful discussions about the observed gamma-ray excess in the Galactic Centre and the modeling of MSP emission. FA acknowledges support by a CIERA postdoctoral fellowship at Northwestern University and the NASA Fermi Grant NNX15AU69G under which this project was initiated. OG acknowledges support from NASA through grant NNX12AG44G and from NSF through grant 1412144.

\end{document}